\documentclass[12pt,preprint]{aastex}
\usepackage{bbm}
\usepackage{mathrsfs}
\usepackage{amssymb,amsmath,float}
\newcommand\al{\alpha}
\newcommand\gam{\gamma}
\newcommand\de{\delta}

\newcommand\lam{\lambda}

\newcommand\ph{\Phi}
\newcommand\vp{\varphi}

\newcommand\om{\omega}

\newcommand\De{\Delta}

\newcommand\<{\langle}
\renewcommand\>{\rangle}

\newcommand\ie{\emph{i.e.}}
\newcommand\eg{\emph{e.g.}}

\newcommand\beq{\begin{equation}}
\newcommand\eeq{\end{equation}}
\newcommand\bea{\begin{eqnarray}}
\newcommand\eea{\end{eqnarray}}
\newcommand\bal{\begin{align}}
\newcommand\eal{\end{align}}

\newcommand\fr{\frac}

\newcommand\half{{\textstyle \frac{1}{2}}}
\newcommand\ap{\approx}

\newcommand\e{\mathrm{e}}

\newcommand\bk{\bold{k}}

\renewcommand\bal{\mbox{\boldmath$\alpha$}}

\newcommand\tn{\tilde{n}}
\newcommand\tm{\tilde{m}}

\newcommand\cT{\delta t}

\newcommand\ct{\tau}

\begin{document}

\title{Measurement of the dispersion of radiation from a steady cosmological source}

\author{Richard Lieu$^1$, Lingze Duan$^1$, and T.W.B. Kibble$^2$}

\affil{$^1$Department of Physics, University of Alabama,
Huntsville, AL 35899\\
$^2$Blackett Laboratory, Imperial College, London SW7~2AZ, U.K.\\}
%$^3$ Space Science Laboratory, University of California, Berkeley, CA, 94720.}

\begin{abstract}
The `missing baryons' of the near universe are believed to be principally in a partially ionized state.  Although passing electromagnetic waves are dispersed by the plasma, the effect has hitherto not been utilized as a means of detection because it is generally believed that a successful observation requires the background source to be highly variable, \ie~the class of sources that could potentially deliver a verdict is limited.  We argue in two stages that this condition is not necessary.  First, by modeling the fluctuations on macroscopic scales as interference between wave packets we show that, in accordance with the ideas advanced by Einstein in 1917, both the behavior of photons as bosons (\ie~the intensity variance has contributions from Poisson and phase noise) and the van-Cittert-Zernike theorem are a consequence of wave-particle duality.  Nevertheless, we then point out that in general the variance on some macroscopic timescale $\tau$ consists of (a) a main contributing term $\propto 1/\tau$, plus (b) a small negative term $\propto 1/\tau^2$ due to the finite size of the wave packets.  If the radiation passes through a dispersive medium, this size will
be enlarged well beyond its vacuum minimum value of $\De t \ap 1/\De\nu$, leading to a more negative (b) term (while (a) remains unchanged) and hence a suppression of the variance w.r.t. the vacuum scenario.  The phenomenon, which is typically at the few parts in 10$^5$ level, enables one to measure cosmological dispersion in principle.  Signal-to-noise estimates, along with systematic issues and how to overcome them, will be presented.

%the total line-of-sight plasma column to a steady cosmological source, such as a quasar.
\end{abstract}

\section{Introduction}

The whereabouts of the baryons in the low redshift universe remains one of the major unsolved problems of cosmology.  While at higher redshifts of $z \gtrsim 2$, observation of the Lyman-$\al$ Forest and the Gunn-Peterson trough (\cite{rau98,bec01}) reveal an adequate amount of baryons, the marked $\sim 50$ \% deficit at $z\lesssim 1$ between the observed density and that expected from nucleosynthesis and the standard cosmological model (\cite{bur98,ade13,kom11}) was pointed out before by \cite{fuk98}.  Contemporaneously, two other papers, \cite{cen99}, and \cite{dav01}, conjectured by cosmological hydrodynamic simulations that the missing baryons in question may have taken refuge in the temperature and density regime of $10^5 - 10^7$ K and $10^{-4} - 10^{-6}$~cm$^{-3}$, as `warm' intergalactic gas in this part of parameter space is hard to trace for many reasons (such as Galactic absorption of the emitted radiation, or the gas may have an unusual composition or ionization state).

Nevertheless, because the models predicted a convergence of warm `filaments' onto the outskirts of clusters of galaxies, the largest bound systems in the universe that were formed
from the collapse of primordial density fluctuations (\cite{pre74,whi78}), there has been an intensified search campaign with nearby clusters as the focal point.  The bulk of the X-ray and Sunyaev-Zeldovich measurements to date (\cite{vik06}, \cite{afs07}, \cite{arn07}, \cite{sun09}, and \cite{ett09}), with the notable exception of \cite{lan13} and \cite{sim11}, reveal that the mass ratio of baryons-to-total matter falls {\it below} the value predicted by the standard model, thereby providing an `inside out' perspective to the existence of baryons beyond the cluster's virial radius.  In addition to being `warm', most recently hints that dark matter is self-interacting, and hence exhibiting `baryonic' characteristics, are pointed out by \eg~\cite{zav13}.  One way or the another, it is clear that the `missing baryon' syndrome persists, and the evidence uncovered has been circumstantial at best, with the only report of a direct detection thus far being the soft X-ray and EUV `excess' emission at $\sim 10^6$~K from cluster halos (\cite{mit98} and \cite{bon02}).

Turning to the question of observational techniques, since most of the baryons at $z \lesssim 1$ (and even higher $z$) are partially  ionized, any electromagnetic radiation passing through the intergalactic medium is expected to be dispersed by them, to a varying degree depending upon the radiation frequency.  Indeed the measurement of dispersion of radiation from a distant source would provide the most reliable means of determining the line-of-sight column density of {\it all} ionized gas components.  However, it is generally thought (but not proven) that for the method to succeed there needs to be a `benchmark', or reference, signal to enable one to time the differential dispersive delay of the various Fourier components constituting the distant emission, and this will exist only if the source is periodic and fast, \ie~a pulsar.

Indeed, while pulsars are a powerful tool for mapping the distribution of Galactic baryons, the fact that they have hitherto not been detected at cosmological separations means nothing can be done to scales beyond the Local Group.  In fact, on such distance scales the majority of the sources seen are quasars, which are for this purpose steady sources.  It should be noted that the very exciting results on fast radio bursts of possible extragalactic origin, {\it viz.} the confirmation by \cite{tho13} of the earlier findings of \cite{lor07}, may represent a change of the observational situation.  Nevertheless, the absence of redshifts from this new class of sources renders it difficult to convert dispersive column densities of their intervening intergalactic media (IGM) into average volume densities out to some specific depths, see \cite{mcq13}.  Even if a means of ascertaining distances is in hand, a detectability analysis by \cite{lor13} indicates that such sources may only be seen out to redshifts of $z \leq 0.5$ by surveys below 1~GHz.  There is also the ambiguity introduced by intrinsic dispersion at the source itself, as pointed out by \cite{mcq13}, which is not as yet well understood for radio transients.  Quasars, on the other hand, have extensively been studied from this viewpoint.  As can be seen in \eg~Figure 1 of \cite{elv02}, $\ap$ 40 -- 45 \% of AGNs show no evidence of warm ionizing absorbers.  Since from \cite{mck07} the lowest detectable ionizing (HII) column of warm absorbers is $\ap 10^{20}$~cm$^{-2}$ which is well below typical IGM values even to nearby quasars, this means a good fraction of quasars can be reliable probes of the IGM baryons {\it if} there is a way of inferring line-of-sight dispersion to such relatively steady sources.

%Thus, while it can easily be shown that intergalactic baryons docause frequency-dependent delay on the timescale of minutes to hours in the arrival of $\sim$ 100 MHz emission from distant quasars,
%quasars with so rapid a variability as to be available
%for this test are those very ones affected by plasma scintillations
%in our local interstellar medium, \eg~\cite{den02}.  On a similar vein, angular broadening of %quasars caused by intergalactic
%scintillation was assessed by Lazio et al. (2008), who concluded
%that the only way of securing a useful observational limit is
%if a quasar nucleus is found to `twinkle' at a
%position close to that of a local pulsar, as data about the latter may then be used to calibrate out the interstellar effects of the Galaxy.

In spite of all the above, we argued in a recent paper (\cite{lie13}) that even the electromagnetic radiation of a {\it steady} source could carry imprint signatures of dispersion, in a way that enables one to recover the total line-of-sight column density of the ionized baryons also.  The imprint is embedded, and can be uncovered only after a careful analysis of the {\it statistical} properties of the fluctuations in the arriving radiation, which are slightly different from normal undispersed radiation.  In this paper we intend to discuss in detail the difference, and how one might detect it.

The premise of our work is the observation of \cite{ein17}, developed subsequently by \cite{pur56}, \cite{han57}, and \cite{man58}, that the variance of photon number fluctuations comprises Poisson and phase noise components, due respectively to the particle and wave nature of radiation.  Indeed as explained in \cite{lie13} the wave packet approach to radiation can account for not only the coherence length, or spatial extent, of interference fringe patterns ({\it viz.} the van Cittert-Zernike theorem, see~\eg~section 10.4.2 of \cite{bor70}) even in the single photon limit, but also why this length is not affected by dispersion.  Here we will show by modeling photons semi-classically as wave packets that the fluctuations of chaotic light from a steady source are fully consistent with Bose-Einstein statistics, \ie~our model is robust as it yields the same results as those from a formal quantum description of light.  In the course of the analysis, it will become apparent that there exists a higher order distortion of the variance, due to the finite size of the wave packet.  The effect is more severe when dispersion by the medium of propagation stretches this size, and herein lies the imprint of the medium.  Since we ignored quantum vacuum corrections, it needs to be emphasized that the approach we adopted is valid in the classical domain when the time and length scales of interest far exceed the wave packet.

%It is clear that the effects of dispersion between the source and observer can be detected when the source comprises a single pulse or a regular sequence of pulses.  The question addressed here is whether the same is true for continuous sources, such as quasars.

\section{The source}

Consider first the radiation at the source, assumed to be at $x=0$.  Let us suppose that it comprises a sequence of wave packets, each of the form
 \beq \ph(t,0)=\sum_j g(t-t_j)e^{i\vp_j},\label{phi} \eeq
where the emission times $t_j$ are randomly distributed with a constant rate of $\lam$ pulses per unit time, and the phases $\vp_j$ randomly distributed between $0$ and $2\pi$.  Here $\lam$ may also be interpreted as the number of photons arriving at the detector per unit time.  In other words, during a long interval $t_{\rm exp}$ of exposure to the source, the expected number of photons entering the detector is $\lam t_{\rm exp}$.  However, there is no one-to-one correspondence between the two sides, ie~one cannot say which of the photons emitted corresponds to which of those detected.

Next, we can reasonably assume that $\int dt \,g(t)=0$.    Let us define
 \beq G(t)=\int dt'\,g(t')g^*(t-t'), \eeq
or equivalently in terms of Fourier transforms
\beq \tilde G(\om) = \int dt \, G(t) e^{i\om t} = |g(\om)|^2. \label{FTG}\eeq
Since the rate $\lam$ is uniform, averages such as $\<\ph(t)\ph^*(t')\>$ or $\<I(t)I(t')\>$ where $I(t)=|\ph(t)|^2$ are functions only of the time difference $t-t'$.  Explicitly, because the emission times $t_j$ are uncorrelated, we need only consider correlations of each wave packet with itself, so
 \beq \<\ph(t,0)\ph^*(t',0)\>=\lam G(t-t'). \label{uncorr} \eeq
That is equivalent to saying that the Fourier components of $\ph$ with different frequencies are uncorrelated:
\beq \<\tilde\ph(\om)\tilde\ph^*(\om')\>=2\pi\lam\de(\om-\om')\tilde G(\om), \eeq where $\<\ X \>$ refers to the average value of $X$ over a time period much longer than the duration of the wave packets.
In particular, the relation
\beq \<I(t)\>=\lam G(0) \label{Iav}\eeq is obtained by setting $t=t'$.

We turn to the intensity correlation function.  Here we have a product of four fields, each of which is a sum, as in (\ref{phi}).  Any contribution where one $t_j$ appears in only a single factor will always vanish.  There are therefore two distinct types of contribution we need to consider.  First we have those from pairwise correlations in which two photons $j,k$ ($j\ne k$) contribute.  In the limit of large $\lam$ (i.e., $\lam\gg \de\nu$, the width of the distribution $\tilde G$) where there is strong overlap between wave packets, this is the dominant contribution.  It is
 \beq \<I(t)I(t')\>=\lam^2[|G(0)|^2+|G(t-t')|^2], \label{IIav1}\eeq
or equivalently,
 \beq \<I(t)I(t')\>-\<I\>^2=\lam^2 |G(t-t')|^2. \label{IIav2}\eeq
So in the large-$\lam$ limit, we have
 \beq \<I^2(t)\>=2\lam^2|G(0)|^2=2\<I\>^2, \label{Ivar}\eeq and the
standard deviation in the intensity equals the mean intensity
(if the source is exposed to the observer for a total time $t_{\rm exp}$, so that the total number of photons is some large number $\lam t_{\rm exp}$, then strictly speaking the omission of the $j=k$ term would mean replacing $\lam^2 t_{\rm exp}^2$ by $\lam t_{\rm exp} (\lam t_{\rm exp}-1)$, but that correction is negligible if $\lam t_{\rm exp} \gg 1$).  Note also that the mean value $\<\ph(t)\ph(t')\>$ will vanish when the random phases are taken into account.

The second type of contribution comes from the case $j=k$ where all four wave packets are the same.  That term is the dominant one in the opposite limit of small $\lam$.  It is
 \beq \<I(t)I(t')\>\ap\lam\int dt''\,|g(t+t'')|^2|g(t'+t'')|^2.
 \label{IIav-wk}\eeq
In either case the correlation function depends only on the time difference, and will tend to zero when $t-t'$ is much larger than the width of the function $G(t)$ (or of $g(t)$).

\section{A distant observer and Gaussian wave packets}

We treat the propagation of radiation from an unresolvable point source to a small telescope as principally a one-dimensional problem $\bk = (k,0,0)$, by ignoring the dynamics of the wave packet in the $y$ and $z$ directions.  Although dispersion broadens the wave packet predominantly along $x$ another effect -- scattering -- does so in all directions.  However, as will be discussed in the end of section 4, for the purpose of this paper scattering is negligible in the radio and even more so at shorter wavelengths.

Let the observer be located at some distant point $x$.  Clearly for a uniform medium of propagation
 \beq \ph(t,x)=\int\fr{d\om}{2\pi} \tilde\ph(\om)e^{-i\om t+ik(\om)x}.\eeq
This is equivalent to saying that the photon pulse shape function $\tilde g(\om)$ is modified by the effect of dispersion:
 \beq \tilde g(\om) \to \tilde g_x(\om)=\tilde g(\om) e^{i[k(\om)-\om/c]x}, \eeq
where the final factor allows for the time development in the absence of dispersion.  Note that from (\ref{FTG}), this means that $G(t)$ is unaffected by dispersion.  It immediately follows that there is no change in the correlation function of $\ph$, given by (\ref{uncorr}), Hence the coherence length of the radiation, defined as the spatial extent of the interference pattern and {\it not} the size of the wave packet (see the end of section 1), is invariant w.r.t. dispersion, as is the large-$\lam$ contribution to the intensity correlation function given by (\ref{IIav1}) or (\ref{IIav2}).  For more elaboration on the coherence length and its invariance, see the end of section 2 of \cite{lie13}.  We shall return to the effect of dispersion on the small-$\lam$ contribution (\ref{IIav-wk}) below.

To proceed further it is convenient to choose a particular form for the function $g$.  Specifically we shall assume that it is a monochromatic beam modulated by a Gaussian envelope function, i.e. a superposition of Fourier components spanning the frequency range $\de\nu$ at $\nu$:
 \beq g(t)=ae^{-i\om_0 t}e^{-t^2/2(\cT)^2}, \eeq
where $a$, $\om_0 = 2\pi \nu$, and \beq \cT = \fr{1}{\sqrt{2}\de\om}= \fr{1}{2\sqrt{2} \pi \de\nu} \label{band} \eeq are constants.  It follows that
 \beq \tilde g(\om) =\sqrt{2\pi}a \cT~e^{-[\cT (\om-\om_0)]^2/2}. \eeq
In this way, a detected photon of chaotic light is treated classically as a pulse emitted by the source, that lasts $\sim$ the reciprocal of the larger of the two bandwidths, the receiver's and the source's.  Dispersive propagation does not alter the pulse spectrum, but can lengthen the pulse duration after emission.
The fact that we took $1/\delta\nu$ as the intrinsic pulse width, where $\delta\nu$ is the {\it receiver's} bandwidth, means the source's emission is presumed to span a larger range of frequencies than allowed by the observer's filter.  In the radio this seems to be reasonable, apart from possibly coherent emission from compact sources like pulsars (\cite{cor76}), since our proposed bandwidth is narrow ($d\nu/\nu \sim$ a few $\times~10^{-5}$, see (\ref{xi})) and radio sources tend to have continuum spectra.  The same is true in the optical provided one avoids the very narrow, $d\nu/\nu <$ 20 \%, emission lines, because our recommended $d\nu/\nu$ is $\ap$ 1/6 from the words below (\ref{III}).

After dispersion, this becomes
 \beq \tilde g_x(\om) =\sqrt{2\pi}a \cT~e^{-[\cT (\om-\om_0)]^2/2}
 e^{i[k(\om)-\om/c]x}. \eeq
Let us assume that the dispersion effect is only slightly nonlinear, i.e., $k(\om)$ in the neighborhood of the peak frequency $\om_0$ is approximately a quadratic function\footnote{The approximation is correct in the context of plasma dispersion provided $\de\nu \ll \nu$, see the discussion before eq. (3) of \cite{lie13}).} of $\om$:
 \beq k(\om)=\fr{\om}{c}+\half\beta(\om-\om_0)^2. \eeq
Note that $c$ here does not have to be the speed of light in vacuum.  It should represent the velocity of a monochromatic wave of frequency $\om_0$ in the medium of interest.  Thus $\tilde g$ becomes
 \beq \tilde g_x(\om) =\sqrt{2\pi}a \cT~\exp {-\half [(\cT)^2-i\beta x](\om-\om_0)^2}. \eeq
Since the effect of dispersion is purely in a phase factor, it does not affect the mean value of the intensity, which is
 \beq \<I\> = \sqrt{\pi}\lam \cT |a|^2. \label{Imean} \eeq

Turning to the intensity correlation function.  Including both the $\lam^2 (\cT)^2$ and $\lam \cT$ terms of (\ref{IIav2}) and (\ref{IIav-wk}), it is
 \bea \<I(t)I(0)\>-\<I\>^2 &=& \lam^2 |a|^4\bigg|\cT~\int d\om\, e^{-i\om t}
 e^{-[\cT (\om-\om_0)]^2} \bigg|^2 + \fr{\lam |a|^4}{1+\xi^2} \int dt' e^{-[(t'+t)^2 + t'^2]/[(\cT)^2 (1+\xi^2)]} \notag\\
 &=& \<I\>^2 ~\left\{\e^{-t^2/[2 (\cT)^2]} + \fr{1}{\lambda \delta t\sqrt{2\pi (1+\xi^2)}}~e^{-t^2/[2(\cT)^2 (1+\xi^2)]}\right\}, \label{ItI0} \eea
where the dispersion stretch factor $\xi = \beta x/(\cT)^2$ may be written in the context of a uniformly expanding universe as \beq \xi = 8.89 \left(\fr{\de\nu/\nu}{4.22 \times 10^{-5}}\right)^2 \left(\fr{\nu}{10^9~{\rm Hz}}\right)^{-1} \left(\fr{n_e}{10^{-7}~{\rm cm}^{-3}}\right) \left(\fr{\ell}{1~{\rm Gpc}}\right), \label{xi} \eeq with $\ell$ being the comoving generalization of the propagation distance $x$, and $n_e$ the mean line-of-sight intergalactic plasma density to the source (see section 3 of \cite{lie13}, where it is also shown in a table that, away from the Galactic disk directions and neglecting sources with unexpectedly large intrinsic ionized columns, the intergalactic medium dominates the column density $n_e\ell$ to a quasar).

It should be mentioned that (\ref{ItI0}) was derived in the context of coherent pulses of radio waves ({\it not}  photons) by another author, as equation (12) of \cite{cor76}.  Moreover, in the $t \ll \cT\sqrt{1+\xi^2}$ limit  a simplification was noted by the author in equation (B6) of Appendix B.  When the pulses are as `microscopic' as photons, however, this limit is actually non-classical, because the Heisenberg Uncertainty Principle for finite $\xi$ is $\De\nu \De t \sim \sqrt{1+\xi^2}$.  We shall henceforth use (\ref{ItI0}) to predict observable imprints of dispersion by focussing our attention upon intensities measured over timescales $\gg \cT\sqrt{1+\xi^2}$ where our hitherto classical treatment of radiation applies.
%It should be emphasized that (\ref{ItI0}) is valid only for timescales $t \gg \cT\sqrt{1+\xi^2}$, as quantum effects may become significant once $t$ is smaller than the time uncertainty\footnote{This estimate of $\de t$ stems from the Heisenberg Uncertainty Principle as applied to a dispersed (non-minimal) wave packet, which reads $\de\nu \de t \ap \sqrt{1+\xi^2} > 1$.} of $\de t \ap \cT\sqrt{1+\xi^2}$.  Even when $t \gg \cT\sqrt{1+\xi^2}$, such effects are not abruptly cut off, just that they are much smaller, \ie~they may still play a role if an experiment exceptionally sensitive to their existence is performed.  We shall return to this point in section 4.
There are two limiting scenarios to discuss.  First is a typical radio passband with $\lam\cT \gg 1$, and second is the visible band (or beyond) with $\lam\cT \ll 1$.  In each case, the signal is also mixed with noise in different proportion.

\section{Radio observations}

Thus far we considered a purely {\it signal limited} source detection environment.
In radio telescopes, there exists also a system noise that can be converted to an equivalent photon counting rate of $\mu$ via the {\it gain} of the telescope.  Thus \eg~for Arecibo where the system noise temperature is $\sim$ 35~K and the gain is $G \ap$ 10~K~Jy$^{-1}$  at $\nu =1$ ~GHz (1 Jansky = 10$^{-26}$~W~m$^{-2}$~Hz$^{-1}$), this converts to the flux density of 3.5~Jy , or~3.5 $\times 10^3$~photons~s$^{-1}$~Hz$^{-1}$ over the entire telescope's collecting aperture of 300~m diameter and $\eta = 0.5$ efficiency.  The value of the dimensionless quantity $\mu \cT$ is set by (\ref{band}) at \beq \mu\cT = 210~{\rm at}~G = 10~{\rm K~Jy}^{-1}. \label{back} \eeq
Assuming $\mu \cT \gg 1$ (and $\lam\cT \gg 1$) henceforth, and that the local radiation is not dispersed, (\ref{ItI0}) is modified to \beq \<I(t)I(0)\>-\<I\>^2 = \left[\pi (\lam + \mu)^2 (\cT)^2  + \sqrt{\fr{\pi}{2}} \mu \cT \right]|a|^4~e^{-t^2/[2(\cT)^2]}  + \lam \cT |a|^4 \sqrt{\fr{\pi}{2(1+\xi^2)}}~e^{-t^2/[2(\cT)^2 (1+\xi^2)]}. \label{II} \eeq  In other words, the unwanted noise is replaced here by a local component of background photons, each being identical in its pulse shape to an intrinsic wave packet from the source.

As indicated in the previous section, the observable imprint of dispersion is upon the variance $\sigma_\ct$ of the average intensity over the macroscopic timescale $\ct \gg \cT\sqrt{1+\xi^2}$, which is given by \beq \sigma_\ct^2 = (\de I_\ct)^2 = \fr{2}{\ct^2} \int_0^\ct dt~(\ct - t) [\<I(t)I(0)\> - \bar I^2]. \label{deI} \eeq  It will be shown that $\sigma_\ct$ is modified by dispersion to decreasing degrees as $\ct \to \infty$.

An important confirmation of the well understood nature of photon fluctuations is obtained by ignoring for the moment the background, \ie~ setting $\mu =0$ in (\ref{II}), and substituting the result into (\ref{deI}).  Enlisting (\ref{Imean}), one gets \beq \sigma_\ct^2 = \pi [\sqrt{2\pi}\lam^2 (\cT)^2 + \lam\cT] |a|^4 \fr{\cT}{\ct} - \sqrt{2\pi}[\sqrt{2\pi}\lam^2 (\cT)^2 + \lam\cT\sqrt{1+\xi^2}]|a|^4 \left(\fr{\cT}{\ct}\right)^2. \label{Ivaries} \eeq Dropping the higher order $(\cT)^2/\ct^2$ terms to see if our model reproduces the standard expression for photon noise in vacuum, (\ref{Ivaries}) may be rewritten as \beq \left(\fr{\de I_\ct}{\bar I}\right)^2 = \fr{\bar n_\gamma^2 + \bar n_\gamma}{\sqrt{2\pi}\lam^2 \cT\ct}, \label{fr} \eeq where the mean photon occupation number per mode is defined at the same value for all modes across the narrow band $\de\nu \ll \nu$ as \beq \bar n_\gam = \sqrt{2\pi} \lam \cT. \label{occ} \eeq  Now the mean number of photons $\bar N_\gam (t)$ arriving during the interval $\ct$ is obviously $\bar N_\gamma (\ct) = \lam\ct$, which must also equal the product of $\bar n_\gam$ and the number of modes $N_{\rm mode}$.  (\ref{occ}) therefore leads us to the very reasonable result of \beq N_{\rm mode} = \fr{\ct}{\sqrt{2\pi}\cT}, \label{Nmode} \eeq indicating that the radiation coherence length is indeed $\ap \cT$.  Moreover, since $(\de I_\ct/\bar I)^2 = (\de N_\gamma/\bar N_\gam)^2 = (\de N_\gam)^2/(\lam^2 \ct^2)$, (\ref{fr}) and (\ref{Nmode}) imply that \beq (\de N_\gam)^2 = N_{\rm mode} (\bar n_\gam^2 + \bar n_\gam). \label{Bose} \eeq  Apart from the small correction terms of order $(\cT/\ct)^2$, then, (\ref{Ivaries}) leads to (\ref{Bose}) which is in full agreement with Bose-Einstein statistics.

This demonstrates the validity of the classical treatment presented.  In fact, it has been shown that for ordinary (chaotic) light from a steady source,  the classical and full quantum calculations of statistical correlations give the same results (see \eg~page 1, Chapter 5 of \cite{lou00}; also eq. (103) and the remarks thereafter of \cite{bal03}).

In the `radio' limit of $\bar n_\gam \ap \lam\cT \gg 1$, \beq \left(\fr{\de I_\ct}{\bar I}\right)^2 = \left(\fr{\de N_\gam}{\bar N_\gam}\right)^2 \ap \fr{1}{N_{\rm mode}} = \fr{1}{2\sqrt{\pi}\ct\de\nu}, \label{radio} \eeq or $(\de I_\ct/\bar I)^2 \ap 1/(\ct\de\nu)$, in accordance with the radiometer equation.

The method of detection utilizes the behavior of $\sigma_\ct^2$ at $\ct \gg \cT\sqrt{1+\xi^2}$, when (\ref{deI}) can be evaluated with the aid of (\ref{II}) to become \beq \sigma_\ct^2 = \pi [\sqrt{2\pi}(\mu + \lam)^2 (\cT)^2 + (\mu+\lam)\cT] |a|^4 \fr{\cT}{\ct} - [2\pi (\mu+\lam)^2 (\cT)^2 + \sqrt{2\pi} (\mu + \lam\sqrt{1+\xi^2})\cT]|a|^4 \left(\fr{\cT}{\ct}\right)^2. \label{Ivary} \eeq  The ratio of the dispersive dependent correction term that suppresses the variance (the preceding $\mu\cT |a|^4$ term is usually negligible w.r.t. this term and can in any case be subtracted) to the main term is \beq r = -\fr{\xi\lam}{\pi (\mu+\lam)^2 \ct} = -2.62 \times 10^{-5} \left(\fr{T_{\rm s}}{10~{\rm K}}\right) \left(\fr{T_{\rm s} + T_{\rm sys}}{45~{\rm K}}\right)^{-2} \left(\fr{\xi\cT/\ct}{0.1}\right), \label{r} \eeq  where the default value of $\ct$ is set at $\ct = 10\xi\cT$ with $\xi\gg 1$ as given in (\ref{xi}), and $T_{\rm s} = 10$~K is because we assumed the default strength of the source to be 1~Jy, \ie~$\lam = \mu$ and both are given by (\ref{back}).

We turn to signal-to-noise issues.  Apart from the variance of the intensity on the timescale $\ct$, the other ingredient to a successful measurement here is an accurate determination of the mean intensity\footnote{It is also necessary to know the occupation number of the source and background,~\ie~$\mu\cT$ and $\lam\cT$ which constitute the second term on the right side of (\ref{Ivary}), but this can be determined to the same accuracy as $\<I\>$ provided the passband filter defining $\de\nu$ is stable (which should not be a big issue because neither the spectra of quasars nor the background has sharp features to enable any drift in the passband to cause a correspondingly large change in the occupation numbers).}, so that all the $\xi$-independent terms of (\ref{Ivary}) are known and can be subtracted from the variance to find the residual.  For this purpose, one can assume that in the $\lam\cT \gg 1$ and $\mu\cT \gg 1$ limit the intensity varies according to the leading term of (\ref{Ivary}) as $\sigma_\ct \ap \<I\> \cT/\ct$, which is $\ll \<I\>$ for $\ct \gg \cT$.  Thus the fluctuations may be regarded as normal (Gaussian).  From (\ref{Gausig}) and (\ref{Gaumsig}), the variance is the main source of error, and is accurate to the fractional uncertainty of $\de\sigma_\ct^2/\sigma_\ct^2 = \sqrt{2/N_s}$, or $4.47 \times 10^{-6}$ if $N_s = 10^{11}$.  Such a value of $N_s$ may be obtained by repeated sampling to a total exposure time of $t_{\rm exp} = 10^4$~s, or 3 hours, at intervals of $\ct = 10\xi\cT \ap 0.237$~ms as suggested in (\ref{r}), using also the channel width of $\de\nu = 42.2$~kHz in (\ref{r}) to simultaneously cover a total bandwidth $\De\nu = 0.1$~GHz at the central frequency $\nu =$~1~GHz.  In precise terms, $N_s =t_{\rm exp}\De\nu/(\ct\de\nu) = t_{\rm exp}\De\nu/(10\xi\cT\de\nu) = 10^{11}$ under this scenario.  Note also that since the denominator $10\xi\cT\de\nu \sim (\de\nu)^2$, we have $N_s \sim 1/(\de\nu)^2$ for fixed $t_{\rm exp}$ and $\De\nu$, hence  the quality of a detection is favored by using a {\it narrower} channel $\de\nu$ whilst maintaining the $\xi \gg 1$ criterion of significant dispersion.  Comparing this uncertainty of $\sqrt{2/N_s}$ with the signal strength of (\ref{r}), one sees that $r$ can be detected at the significance of $\ap 5.91\sigma$, increasing to $8.55\sigma$ at the optimal source brightness of 3.5~Jy where the product of the two temperature dependent coefficients in (\ref{r}) is at its maximum.

In fact, using (\ref{r}) the signal-to-noise ratio may be written as \bea \fr{|r|}{\de r} &=& 5.91 \left(\fr{T_{\rm s}}{10~{\rm K}}\right) \left(\fr{T_{\rm s} + T_{\rm sys}}{45~{\rm K}}\right)^{-2} \left(\fr{\xi\cT/\ct}{0.1}\right)^{3/2} \left(\fr{\de\nu}{160~{\rm kHz}}\right)^{-1} \notag\\
&\times&  \left(\fr{\De\nu}{0.1~{\rm GHz}}\right)^{1/2} \left(\fr{\nu}{1~{\rm GHz}}\right)^{3/2} \left(\fr{t_{\rm exp}}{10^4~{\rm s}}\right)^{1/2} \left(\fr{n_e \ell}{10^{-7}~{\rm cm}^{-3}~{\rm Gpc}}\right)^{-1/2}. \label{sn} \eea  One caveat to be mindful of is that the weakness of the signal $r$ means the actual detection significance will be reduced by extra (system) noise, including and especially radio frequency interference and gain variations, which have the effect of increasing the background $\mu$ by typically 10 \%, see \cite{tuc09}.

%Such contaminating components of the data have therefore got to be kept to a minimum by every means.

We should  also discuss the potential complication of scattering by plasma clumps, which can also broaden radio wave packets in all 3 dimensions.  However, for a mean intergalactic plasma column density appropriate to a 1~Gpc source a wave packet is typically broadened to last the duration of $\ap 10^{-6}$~s, (\cite{bha04}).  This is still very small compared with the $\ap 10^{-4}$~s of dispersion effect in the propagation direction $x$, {\it viz.} (\ref{xi}).  We shall therefore neglect it.

Given the formidability of securing a robust detection in the radio, it may be advisable to first perform a feasibility test by observing an extragalactic source at low Galactic latitude $b$ where dispersion by the interstellar medium (ISM) sets a `bottom line' minimum in the plasma column $n_e \ell$ (hence $\xi$) of order \beq n_e^{\rm ISM} \ell > 10^{-7}~{\rm  cm}^{-3}~{\rm Gpc,~for}~\ell \gtrsim 1.5~{\rm kpc}, \label{ism} \eeq (see \eg~Table 1 of \cite{david69}), which according to Table 1 below is comparable to the IGM column, thereby affording one a rough idea of what result to expect.  We suggest looking at a low $b$ quasar such as J2109+353, which has $(l,b) = (80.3, -8.35)$ and a flux of 1.2~Jy at 1.4~GHz, see Table 1 of \cite{im07}.  The reason for avoiding any Galactic sources is that these will inevitably have to be pulsars, except their emission may occur in a highly unusual form as coherent bunches of photons (\cite{cor76}) and therefore the receiver's passband $1/\de\nu$ may underestimate the intrinsic wave packet size and jeopardize the measurement -- see section 6 for further discussions.

The conclusion is that even with the largest radio telescope available observations in this band can at best yield a marginal detection of cosmological dispersion with a set of fortuitous and very fine tuned parameters.  It is hard to see how to develop the technique to become a utility for mapping the baryonic content of the entire near universe.  To achieve that, one must explore other wavelengths.

\section{Optical observations}

In the opposite limit of $\lam\cT \ll 1$ and $\mu\cT \ll 1$, such as optical observation of quasars, (\ref{II}) simplifies to \beq \<I(t)I(0)\>-\<I\>^2  = \sqrt{\fr{\pi}{2}} \mu \cT |a|^4~e^{-t^2/[2(\cT)^2]}  + \lam \cT |a|^4 \sqrt{\fr{\pi}{2(1+\xi^2)}}~e^{-t^2/[2(\cT)^2 (1+\xi^2)]}. \label{III} \eeq For typical observation through a Gaussian V-band filter of $\nu =$~500 THz and $\de\nu/\nu = 1/6$ (or 0.1 $\mu$m passband at 0.6 $\mu$m wavelength) one has, by (\ref{band}) and (\ref{xi}),  \beq \cT\sqrt{1+\xi^2} \ap 3.75 \times 10^{-13}~{\rm s,~at}~\xi = 278 \label{xiT} \eeq
for a source $\ap$~1~Gpc away.

The imprint of dispersion upon the autocorrelation function is not affected at optical wavelengths by the (undesirable) $\lam^2 (\cT)^2$ and $\mu^2 (\cT)^2$ terms, as such terms are negligible in the Poisson limit of photon noise.  There is also the advantage of the generally higher signal-to-background ratio $\lam/\mu$ here (see below).  But the low photon count rate, the narrowness of the visible band, and the smaller value of $\xi$ present this type of observations with its own challenges.

Like the radio band, one may proceed to measure the dispersive column density via the variance of intensity fluctuations on timescales longer than $\cT\sqrt{1+\xi^2}$, by applying (\ref{III}) to (\ref{deI}) to get \beq \sigma_\ct^2 = (\de I_\ct)^2 = \pi (\mu + \lam)\cT |a|^4 \fr{\cT}{\ct} - \sqrt{2\pi} (\mu + \lam\sqrt{1+\xi^2})\cT |a|^4 \left(\fr{\cT}{\ct}\right)^2,~{\rm for}~\ct \gg \cT\sqrt{1+\xi^2}. \label{dIct} \eeq  As already discussed in the last section, the $O(\cT/\ct)$ term spells Poisson statistics in this limit.  To check this again, note that when (\ref{dIct}) is used in conjunction with (\ref{Imean}) this lowest order contribution to the variance satisfies the relation \beq \left(\fr{\de I_\ct}{\bar I}\right)^2 = \left(\fr{\de N_\gam}{\bar N_\gam}\right)^2 = \fr{1}{(\lam + \mu)\ct} = \fr{1}{\bar N_\gam}, \label{Poisson} \eeq independently of dispersion, where $N_\gam$ is the total number of arriving photons in the interval $\ct \gg \cT\sqrt{1+\xi^2}$.  This result is a direct consequence of (\ref{Nmode}) and (\ref{Bose}) in the `optical' limit of $\bar n_\gam \ap (\mu + \lam)\cT \ll 1$ (end of section 2).

Over this same interval $\ct$, however, the imprint of dispersion is upon the {\it modification} to Poisson ($\sigma_\ct^2 \propto 1/\ct$) fluctuations from the very last term of (\ref{dIct}).  Explicitly, the ratio of this term to the rest of $\sigma_\ct^2$ is, assuming $\xi \gg 1$, \beq r = -\sqrt{\fr{2}{\pi}}
\fr{\xi\lam}{\mu + \lam}
\fr{\cT}{\ct} = -3 \times 10^{-5} \left(\fr{\ct}{10~{\rm ns}}\right)^{-1} \left(\fr{\de\nu}{83.3~{\rm THz}}\right) \left(\fr{\nu}{500~{\rm THz}}\right)^{-3} \left(\fr{n_e \ell}{10^{-7}~{\rm cm}^{-3}~{\rm Gpc}}\right), \label{ratio} \eeq which is $\xi$-dependent\footnote{Note that when calculating $r$ we ignored the $\mu(\cT)^3/\ct^2$ term in (\ref{dIct}), since it is much less than the $\mu(\cT)^2/\ct$ term there.} (the last expression is written with $\mu =0$ in mind).  Thus one expects to see a {\it significantly larger} non-Poisson correction to the curve when there is dispersion, signal-to-noise-permitting, in the form of a suppression of the variance $(\de N_\gamma)^2$ of the photon counts from the Poisson value of $(\de N_\gam)^2 = \bar N_\gam$.

\subsection{Photon counting on short timescales}

Our calculation is done in the context of the 10m telescope at the Keck Observatory, although the formulae are presented in such a way as to facilitate adaptation to another environment, and the final results as shown in Table 1 contain sources observed by the Lick 3m as well.  Assuming a quantum efficiency of 16 \% (\cite{ham06}) for the photomultiplier detector, a quasar with V-band (see the beginning of this section for filter specifics) magnitude mV$=$ 15 would deliver $\lam = 4 \times 10^4$ photons s$^{-1}$, against a background of $\mu = 2,800$ photons s$^{-1}$ if we assume the dark sky background  has the surface brightness of 20 mV~arcsec$^{-2}$ and a point spread function of 3" diameter for the quasar.  Thus the data are signal dominated, with $\lam \gg \mu$.

We explore the prospect of using the above parameters for an observing run.  Suppose one of them is exposed\footnote{The exposure time $t_{\rm exp}$ must satisfy the condition $\lam t_{\rm exp} \gg 1$, see section 1.  The inequality ensures that there are many source photons detected during the observation.} for a time $t_{\rm exp}=$~10$^4$~s ($\ap$~3 hours), with the arrival time of individual photons recorded at the time resolution of $\ct =$~10~ns (10$^{-8}$~s, photomultiplier tubes (PMT) with this requisite response time and without dead time issues are available, see below), \ie~there are $N_s = t_{\rm exp}/\ct \ap 10^{12}$ time bins to measure the variance $\sigma_\ct^2$.  Neglecting $\mu$ in (\ref{ratio}), the imprint of dispersion is an excess of the mean photon count per bin from the variance, by the  fractional amount $r = -3 \times 10^{-5}$.  Now from (\ref{ms}) of the Appendix, the random error in this fraction is \beq \de r \ap \sqrt{\fr{2}{N_s}} = 1.4 \times 10^{-6} \left(\fr{\ct}{10~{\rm ns}}\right)^{1/2} \left(\fr{t_{\rm exp}}{10^4~{\rm s}}\right)^{-1/2}, \label{der} \eeq  The signal-to-noise ratio is obtained by combining (\ref{ratio}) with (\ref{der}), as
\bea \fr{|r|}{\de r} &=& 19.8 \left(\fr{\lam}{4\times 10^4~{\rm s}^{-1}}\right) \left(\fr{\lam + \mu}{4.28 \times 10^4~{\rm s}^{-1}}\right)^{-1}  \left(\fr{\ct}{10~{\rm ns}}\right)^{-3/2}\notag\\
&\times& \left(\fr{t_{\rm exp}}{10^4~{\rm s}}\right)^{1/2} \left(\fr{\de\nu}{83.3~{\rm THz}}\right) \left(\fr{\nu}{500~{\rm THz}}\right)^{-3} \left(\fr{n_e \ell}{10^{-7}~{\rm cm}^{-3}~{\rm Gpc}}\right). \label{SNR} \eea  Hence for the prescribed observing conditions one can expect a $\ap 20\sigma$ detection of the dispersion signal, and a tight ensuing constraint on the intervening IGM column density to the quasar via $\xi$.

The reader can also verify readily that under the above scenario the $\bar n_2 \ap 9.16 \times 10^4$ `double photon' bins expected from a pure Poisson distribution is sufficiently large to enable a slight reduction in the number $n_2$ of such bins to produce the corresponding suppression of the variance to a sub-Poisson value beneath the mean. To be precise, the variance of $N_\gam$ will come down from the mean by the fraction of $3 \times 10^{-5}$ if the average number of `double photon' bins is $\bar n_2=8.56 \times 10^4$ instead of $\bar n_2 = 9.16 \times 10^4$.   In another way of understanding why this decrease is 20$\sigma$ significant, one could take the difference between the above two values of $n_2$, and divide it by the statistical fluctuation of $n_2$, {\it viz.}~$\sqrt{\bar n_2} \ap 303$.  More generally, it can be shown without too much difficulty that for $\ct \gg \cT\sqrt{1+\xi^2}$, \beq \bar n_2 = \tfrac{1}{2} (\lam + \mu)^2 \ct^2 N_s - \sqrt{\fr{1+\xi^2}{2\pi}} \lam\cT N_s, \label{n2} \eeq  from which one can see that the slight sub-Poisson behavior due to the last term is enhanced by dispersion.

A sensitivity limit of this technique exists, however, when there are too few arriving photons from a faint source, \ie~if the mean count per bin $\bar m = \lam\ct$ is so low that the first term on the right side of (\ref{n2}) becomes as small as the second.  Since $n_2$ obviously cannot be negative, the dispersion signal is compromised once this limit is reached, and more and more so beyond it.  Explicitly, the smallest variance afforded by any photon counting data of a given mean $\bar m$ occurs when $\bar n_2 = 0$ and $\bar n_1 = \bar m - \bar m^2$, and the variance of the configuration lies below the mean by the fractional amount $(\bar m - \sigma^2)/\bar m = \bar m$.  If $\bar m < |r|$ (where $r$ is given by (\ref{ratio})), the measurement will not be able to recover the full signal.  Thus a faint source will require large $\ct$ to satisfy the $\bar n_2 \gg 1$ criterion for full signal detection, but (\ref{SNR}) asserts that as $\ct$ increases the signal-to-noise ratio deteriorates.  The use of too large a time window $\ct$ would also put one closer to the timescales of atmospheric turbulence effects, which tends to induce super-Poisson variance at the few \% level over durations of 1~ms to 1~s (\cite{tok03}), \ie~although this concern is not imminent it should be kept in mind.  In general, then, a correspondingly longer exposure time $t_{\rm exp}=N_s\ct$, is needed to maintain the signal-to-noise ratio $|r|/\de r$.  Since, for $\bar n_2$  of (\ref{n2}) to stay positive $\ct$ must at least be sufficiently large to enable $\lam\ct$ to match the constant $\ap \sqrt{\xi\lam\cT}$ , {\it viz.} \beq \ct \gg 7.75 \times 10^{-10} \left(\fr{\lam}{4 \times 10^4~{\rm s}^{-1}}\right)^{-1/2} \left(\fr{\de\nu}{83.3~{\rm THz}}\right)^{1/2} \left(\fr{\nu}{500~{\rm THz}}\right)^{-3/2} \left(\fr{n_e \ell}{10^{-7}~{\rm cm}^{-3}~{\rm Gpc}}\right)^{1/2}~{\rm s}, \label{n2c} \eeq one sees from (\ref{SNR}) that to maintain $|r|/\de r$ the exposure time $t_{\rm exp}$ has to be raised for a faint source according to the scaling\footnote{This scaling is to be contrasted with then exposure $t_{\rm exp} \sim 1/\lam$ necessary to detect a source photometrically to a certain level of significance under the same {\it signal limited} condition as here.} $t_{\rm exp} \sim 1/\lam^{3/2}$ (alternatively the telescope aperture has to enlarge by the same scaling), because in (\ref{SNR}) $t_{\rm exp} \sim \ct^3$ for a given $|r|/\de r$ and $\ct_{\rm min} \sim \lam^{-1/2}$ from (\ref{n2c}).  Moreover, eventually the background $\mu$ will also become significant, in which case $t_{\rm exp}$ will have to be even larger, although the fact that faintness usually implies remoteness would work in one's favor because it means $|r|$ is larger as a result of the column $\ell$ being so.

Although the visible regime appears more promising than the radio, it is still advisable to test the effect, which remains rather feeble, using a `calibration' source as described in the end of section 4, where we also explained why a source located 1.5 kpc or more away in the direction of the Galactic disk would suffice.  Specifically one can enlist here a bright disk star (\ie~pulsars are not the only choice even if one observes only sources within the Galaxy), such as P Cygni (34 Cyg) which is 1.8 kpc away (\cite{bal10}) but has mV$= 4.8$ from the \texttt{SIMBAD} Astronomical Database.  Signal-to-noise estimates for a fainter (mV$= 10$) star are given in Table 1, from which it can be seen that P Cygni should be a relatively easy target.

\subsection{An outline of the basic design}

Here we present the elements of a feasible observational scheme. Our intention here is not to define in detail the optimal technique, however.  This is work in progress, to appear in a separate publication.

Our goal is to measure the reduction of $n_2$, which can be done by repeatedly counting the number of photons received by the detector using a fixed counting cycle (or window span). After a sufficient number of photons have been received, a histogram of the per-cycle photon count can be developed and is compared to the Poisson distribution.  The temporal width of the counting cycle has to be chosen in such a way that the expected number of `double photon' cycles is statistically robust while the random noise is kept low to ensure an appropriate signal-to-noise ratio, as already explained.  Note that the latter requirement usually implies that the `triple photon' probability has to be prohibitively low,  which leaves $n_2$ as the only parameter that carries the signature of the sub-Poisson distribution.

Several practical issues associated with photon counters should be taken into account, including dead time, afterpulsing and dark counts. The dead time of a photon counter tends to make a perfect Poisson distribution appear to be sub-Poisson. Such an effect could mask the sub-Poisson distribution due to dispersion if proper care is not taken. For example, the counting circuits of a PMT usually introduce a dead time of a few tens of nanoseconds. Such a photon counter would not be able to capture any "double photon" counts if a 1-10~ns counting cycle is used. To address this problem, we propose to use two identical PMT counters separated by a 50:50 beamsplitter. The outputs of the two counters are combined with a sub-ns timing error to form one single stream of photon counts before being gated by a clock. To simplify the data analysis, we can further make the counting cycle same as the counter dead time (by adjusting the PMT electronics and the clock frequency). Such a detection system in principle can capture all the "single photon" counts and half of the `double photon' counts. The latter is because 50 \% of the photon pairs arriving within a counting cycle fall onto different PMTs and, therefore, can effectively be tallied with a worsening in the signal-to-noise by the factor of $\sqrt{2}$.

On the other hand, afterpulsing and dark counts tend to make a perfect Poisson distribution appear to be super-Poisson by adding artificial multi-photon counts. However, with the help of an ideal Poisson source, both effects can be fairly well calibrated and subsequently taken out of the dispersion measurement. Recent development of hybrid photodetectors also offers a detector with almost zero afterpulsing (\cite{suy08}). The counter calibration can be done off-line in the lab. Alternatively, the task can also be done on site by pointing the telescope first to a nearby star and attenuate the incoming photon intensity to a level comparable to the intended quasar.

In Table 1 we show the typical plasma column density $n_e \ell$ to various astrophysical sources, the first of which is non-cosmological (a nearby star), and the recommended parameters at the Lick or Keck Observatory (the former restricted to stars and the brightest quasar 3C273 as test beds) to measure $n_e \ell$ via $\xi$.

%\newpage

\begin{table}
%\vspace{-0.7cm}
    \begin{tabular}{|l|l|l|l|l|l|l|}
        \hline
        mV  & Obs. & $\lam$  & Column $n_e \ell$  & $\ct$  & Statistical & $\de n_2/n_2$ \\ \
        of source & & (in s$^{-1}$)  & (in cm$^{-3}$~Gpc) & (in ns) & significance & \\ \hline
        10 (star) & Lick 3m & $4 \times 10^6$ & $10^{-8}$ & 1 & $47.2\sigma$ & $6\times 10^5/8\times 10^7$ \\
        13 (3C273) & Lick 3m & $2.5 \times 10^4$ & $10^{-7}$ & 10 & $13.5\sigma$ & $3,750/38,642$ \\
        15 (quasar) & Keck 10m & $4 \times 10^4$ & $10^{-7}$ & 10 & $14.0\sigma$ & $6,000/91,592$ \\
        18 (quasar) & Keck 10m & 2,560 & $3 \times 10^{-7}$ & 20 & 7.59$\sigma$ & $576/2,873$ \\
        \hline
    \end{tabular}
\vspace{0.5cm}
    \caption{Observation of dispersion induced `photon anti-bunching' to determine the line-of-sight plasma column density to astrophysical sources in the V band.  In each case the night sky background is assumed to be 2,800 photons s$^{-1}$, and the exposure time is 10$^4$~s.  The last column gives the $\de n_2$ {\it reduction} in the number of double photon time bins (of width $\ct$) due to dispersion, and the expected number $n_2$ of such bins in the limit of no dispersion. The statistical significance is computed with the use of a beam splitter in mind to counter dead time problem (see section 5.2), \ie~the number of $\sigma$ is smaller than $\de n_2/\sqrt{n_2}$ by a factor of $\sqrt{2}$ as a result.}
\end{table}

\section{Conclusion}

In respect of the quest for the missing baryons in the near universe, an effort at the forefront of contemporary cosmology, we proposed and developed a new technique that complements the current reliance on fast radio transients.  By examining the detailed  statistical fluctuations of the electromagnetic radiation arriving from distant quasars, we showed that it is possible to detect the imprint of dispersion and infer the total line-of-sight column density of partially ionized (\ie~ the bulk of the missing) baryons to the source.

The technique as applied to the radio and visible bands are discussed.  For each case it is found that different challenges exist.  In the radio, systematic noises like the broad band interference needs to be kept to a minimum, and random errors reduced by adequate exposure using simultaneously many frequency channels.  In the visible the background is less of a problem, rather the scarcity of photons and the relatively smaller effect of dispersion.
However, all these issues can potentially be overcome, and the prospect of success is reachable.  On balance, the visible band appears more promising, and could be developed into a widely applicable utility for mapping the low $z$ plasma columns across the sky.

The authors acknowledge Eric Korpela, Michael Lampton, and Stuart Bowyer for discussions.

\appendix

\section{Sample variance distribution; correlation between sample mean and variance}

For $N_s$ samples (each is in the context of this paper a measurement
of either radio intensity or optical photon counts over some small
interval of time) of a variate $n$ drawn from a parent population of
central moments ${\mu_m, m=1,2,\cdots}$,~{\it viz.}
\beq \mu_j = \<(m - \mu)^j \> = \fr{1}{N_s} \sum_{i=1}^{N_s} (n_i -
m)^j, \label{mus} \eeq  where $\mu = \<m\>$ is the true mean (not to be confused with the background count rate $\mu$ of section 4), and the sample mean and variance are
denoted by $m$ and $\sigma^2$ respectively, and are defined as \beq m=\fr{1}{N_s} \sum_{i=1}^{N_s} n_i;~\sigma^2 = \left(\fr{1}{N_s} \sum_{i=1}^{N_s} n_i^2\right) - m^2. \label{mvar} \eeq
The expectation value for the sample
variance, $\<\sigma^2\> = (N_s -1)\mu_2/N_s$ is $\ap\mu_2$ to a
fractional error $\ap 1/N_s$, and that of the variance of the sample
variance is \beq \<(\de\sigma^2)^2\> = \fr{(N_s -1)^2}{N_s^3}\mu_4 -
\fr{(N_s -1)(N_s -3)}{N_s^3}\mu_2^2, \eeq which also simplifies to
\beq \<(\de\sigma^2)^2\> = \fr{\mu_4 - \mu_2^2}{N_s}, \label{dsig2}
\eeq again correct to a fractional error $\ap 1/N_s$.

Let us now apply (\ref{dsig2}).  For a normal distribution where $\mu_2 = \sigma^2$ and $\mu_4 = 3\sigma^4$, it implies, adopting slightly loose notations, \beq \fr{\de\sigma^2}{\<\sigma^2\>} = \sqrt{\fr{2}{N_s}}. \label{Gausig} \eeq  This is to be contrasted with the standard expression for the uncertainty in the mean, \beq \fr{\de m}{\<m\>} = \fr{\sigma}{\sqrt{N_s}}, \label{Gaumsig} \eeq which is much smaller in the limit $\sigma \ll \mu$.

For Poisson fluctuations the variance
$\mu_2 = \sigma^2 =  \mu$ and $\mu_4 = \mu (1+3\mu)$, and it
readily follows from (\ref{dsig2}) that the relative uncertainty $\de\sigma^2/\sigma^2$ in
the sample variance $\sigma^2$ is
\beq \fr{\de\sigma^2}{\<\sigma^2\>} =
 \fr{1}{\sqrt{N_s \mu}}, \label{Poisig} \eeq and correct to
the relative
accuracy $\ap 1/N_s$.  (\ref{Poisig}) is the same as the relative uncertainty in the sample mean, $\de m /\<m\>$, and this peculiar feature is specific to the $\mu \ll 1$ limit only.  In the opposite limit, the sample variance fluctuates much more than the mean.

For completeness even though the next result is not used in the paper, we include phase-noise fluctuations of radio observations, \ie~the $\mu \gg
 1$ limit of Bose-Einstein statistics where $\mu_2 = \mu^2$ and $\mu_4 = \mu (1+\mu)(1+9\mu + 9\mu^2) \ap 9 \mu^4$, one obtains instead the relation \beq \fr{\de\sigma^2}{\<\sigma^2\>}
 = \sqrt{\fr{8}{N_s}} \label{BEsig} \eeq for the relative uncertainty in $\<\sigma^2\>$ from (\ref{dsig2}).

We are also interested in the relative uncertainty in the {\it difference} between the sample mean and variance, {\it viz.}~$\de (m - \sigma^2)/\sigma^2$ of a Poisson distribution.  To begin, let us express the variance of $x=m - \sigma^2$ as \beq \<(\de x)^2\> = \<(\de m)^2\> + \<(\de\sigma^2)^2\> - 2{\rm cov}(m, \sigma^2), \label{tandem} \eeq where the covariance function of two variates $y$ and $z$ is defined in the usual manner as $$ {\rm cov}(y,z) = \<yz\> - \<y\> \<z\>. $$  In this case, since the Poisson $\<m\> = \<\sigma^2\> = \mu$ we may write $$ {\rm cov}(m, \sigma^2) = \<m \sigma^2\> - \mu^2. $$  This equation may be recast in terms of the new variates $\tn_i = n_i - \mu$ and $\tm = m - \mu$, as \beq {\rm cov}(m,\sigma^2) = \<\tm\sigma^2\> = \<\tn_1 \sigma^2\> = \fr{1}{N_s}\<\tn_1 w\>, \label{cov} \eeq where $$ w = \sum_{i=1}^{N_s} (\tn_i - \tm)^2.$$  Now $w$ is a sum of the products $\tn_i^2$ and $\tn_i \tn_j$ for $i\neq j$.  Most of these products actually do not affect the calculation of $\<\tn_1 w\>$, since \eg~$\<\tn_1\tn_i\tn_j\> = 0$ for $i \neq j$ and $\<\tn_1 \tn_i^2\> = 0$ for $i\neq 1$.

The quantity $\<\tn_1 w\>$ may therefore be simplified to read $\<\tn_1 w\> = \al \<\tn_1^3\>$ where $\al = 1$ is accurate to the relative error of $1/N_s$ and $\<\tn_1^3\> = \mu_3 =\mu$ is the 3rd moment of the Poisson distribution.  Substituting into (\ref{cov}), we obtain \beq {\rm cov}(m,\sigma^2) = \fr{\mu}{N_s}. \label{covans}  \eeq  Since $\<(\de m)^2\> = \<\sigma^2\>/N_s = \mu/N_s$ and $\<(\de\sigma^2)^2\> = (\mu + 2\mu^2)/N_s$ from (\ref{dsig2}) and the fact that the Poisson $\mu_4 = \mu + 3\mu^2$ and $\mu_2 = \mu$, these results may be applied along with (\ref{covans}) to (\ref{tandem}) to arrive at \beq \<(\de x)^2\> = \fr{2\mu^2}{N_s}. \label{ms} \eeq  In the limit of small counts per bin, \ie~$n \ap \mu \ll 1$, the variance $\<(\de x)^2\>$ in the {\it difference} between the sample mean and variance of a Poisson distribution, $n - \sigma^2$ is {\it much smaller} than the $\mu/N_s$ variance in the sample mean and variance themselves.

\end{document}